\documentclass[
reprint,longbibliography,
superscriptaddress,
amsmath,amssymb,
aps,
prl,
]{revtex4-2}

\usepackage{graphicx}
\usepackage{amsmath}
\usepackage{color}
\usepackage[normalem]{ulem}
\usepackage[colorlinks, linkcolor= blue, citecolor = blue, urlcolor=blue]{hyperref}

\begin{document}

\title{Short-Pulsed Metamaterials}
\author{Carlo Rizza} \email{carlo.rizza@univaq.it}
\affiliation{Department of Physical and Chemical Sciences, University of L'Aquila, Via Vetoio 1, I-67100 L'Aquila, Italy}
\author{Giuseppe Castaldi}
\affiliation{University of Sannio, Department of Engineering, Fields \& Waves Lab, Benevento, I-82100, Italy} 
\author{Vincenzo Galdi} \email{vgaldi@unisannio.it}
\affiliation{University of Sannio, Department of Engineering, Fields \& Waves Lab, Benevento, I-82100, Italy}

\begin{abstract}
We study a class of temporal metamaterials characterized by time-varying dielectric permittivity waveforms of duration much smaller than the characteristic wave-dynamical timescale. In the analogy between spatial and temporal metamaterials, such 
a {\em short-pulsed} regime can be viewed as the temporal counterpart of metasurfaces.
We introduce a general and compact analytical formalism for modeling the interaction of a short-pulsed metamaterial with an electromagnetic wavepacket. Specifically, we elucidate the role of local and nonlocal effects, as well as of the time-reversal symmetry breaking, and we show how they can be harnessed to perform elementary analog computing, such as first and second derivatives. Our theory, validated against full-wave numerical simulations,  suggests a novel route for manipulating electromagnetic waves without relying on long, periodic temporal modulations. Just as metasurfaces have played a pivotal role in the technological viability and practical applicability of conventional (spatial) metamaterials, short-pulsed metamaterials may catalyze the development of temporal and space-time metamaterials. 
\end{abstract}

\maketitle

Conventional metamaterials and metasurfaces are artificial structures relying on suitably designed {\em spatial} arrangements (volumetric and planar, respectively) of deeply sub-wavelength meta-atoms that are engineered so as to attain desired physical responses not necessarily found in natural materials \cite{Capolino:2009ta}. Currently, the growing availability of reconfigurable meta-atoms, whose response can be rapidly changed in time, has granted access to the {\em temporal} dimension as well, and 
has revamped the interest in studying  wave interactions with time-varying media \cite{Morgenthaler:1958vm,Oliner:1961wp,Fante:1971to}, within the emerging frameworks of ``temporal''  and ``space-time'' metamaterials \cite{Caloz:2020sm1,Caloz:2020sm2,Engheta:2021mw}.

With specific reference to optics and photonics, by relying on the space-time duality, a variety of concepts and tools typically utilized in spatially variant configurations have been translated to time-varying scenarios, starting from simple counterparts such as temporal boundaries \cite{Xiao:2014ra} and slabs \cite{Mendonca:2003tb,Ramaccia:2020lp}, and moving to more sophisticated aspects such as Brewster angle \cite{Pacheco:2021te}, Faraday rotation \cite{Li:2022na}, diffraction gratings \cite{Taravati:2019gs,Galiffi:2020wa}, effective-medium theory  \cite{Pacheco:2020em,Huidobro:2021ht} and related nonlocal corrections \cite{Torrent:2020ss,Rizza:2022ne}, transfer-matrix modeling \cite{Li:2019tm,Xu:2022gt}, impedance transformers \cite{Pacheco:2020at,Castaldi:2021es} and
filters \cite{Ramaccia:2021tm}, just to mention a few. The reader is referred to \cite{Galiffi:2022po} (and references therein) for a recent comprehensive review.

\begin{figure}[t]
	\centering
	\includegraphics[width=.95\linewidth]{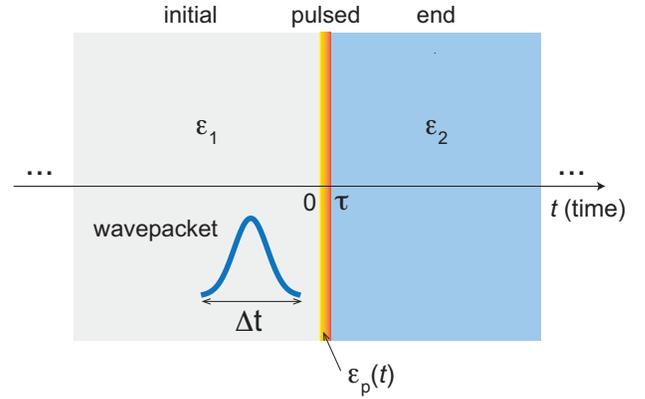}
	\caption{Schematic of an SPM (details in the text).}
	\label{Fig1}
\end{figure}

Typical temporal metamaterials, based on periodic temporal modulations of the dielectric permittivity, can be viewed as the analog of spatial, {\em volumetric} multilayered metamaterials and, under appropriate conditions, can be likewise homogenized \cite{Pacheco:2020em}. Looking at the evolution of conventional (spatial) metamaterials, the surge of {\em lower-dimensional} implementations (metasurfaces) has played a pivotal role in determining the technological viability and improving the efficiency and integrability in scenarios of practical interest, and arguably constitutes one of the most promising research directions \cite{Quevedo_Teruel:2019ro}. Based on the space-time analogy mentioned above, one might wonder to what extent a similar path may be followed in the temporal case as well; this implies moving from {\em periodic} temporal modulations to
{\em pulsed} variations much shorter than the characteristic wave-dynamical timescale. 

Although some general scenarios of temporally resolved modulations have been studied \cite{Mendonca:2003tb,Ramaccia:2020lp},  such a {\em short-pulsed metamaterial} (SPM) regime  remains largely unexplored, and its systematic investigation motivates our study here. 
To this aim, we develop a general theory to model the interaction of an electromagnetic (EM) wavepacket with an arbitrary SPM. Via a multiscale asymptotic analysis, we derive a compact formalism that elucidates the role of the time-reversal symmetry breaking and local/nonlocal effects in the SPM regime. In addition, we show how these effects can be harnessed to synthesize some elementary analog-computing functionalities. 

As schematically illustrated in Fig. \ref{Fig1}, we consider an EM wavepacket (with characteristic timescale $\Delta t$) propagating in a spatially unbounded, nonmagnetic, temporal metamaterial, described by a time-varying relative dielectric permittivity 
\begin{eqnarray}
\label{temp}
\varepsilon(t)=
\left \{ \begin{array}{ll}
 \varepsilon_1, \quad t< 0, \\
  \varepsilon_p(t), \quad 0 <t <\tau, \\
\varepsilon_2, \quad t > \tau,
\end{array}
\right.
\end{eqnarray}
where $\varepsilon_1$ and $\varepsilon_2$ are constant in time, 
whereas $\varepsilon_p(t)$ is a pulsed waveform of duration $\tau\ll\Delta t$, consistently with our SPM definition. As in previous studies on temporal metamaterials \cite{Pacheco:2020em}, we assume to operate far away from any material resonance frequencies, and therefore we neglect temporal dispersion \cite{Morgenthaler:1958vm,Solis:2021fa}. Moreover, we generally assume that the EM field can experience a so-called ``temporal boundary'' \cite{Xiao:2014ra}, at which the dielectric permittivity (uniform in space) changes abruptly everywhere at a specific time instant $t_b$. Specifically, we consider a discontinuous behavior at $t_b=0,\tau$; while this idealization is clearly unphysical, our results remain valid as long as the rising/falling times are much shorter than the duration $\tau$ (see further details on full-wave simulations in \cite{SuppMat}). The  dynamics of the electric induction field $\bf D$ is governed by the vector wave equation  $\partial_{tt}^2 {\bf D}  - c^2 \varepsilon^{-1}(t) \nabla^2 {\bf D} =0$, with $c$ denoting the wave speed in vacuum. In analogy with conventional spatially modulated metamaterials (where the inclusions size are much smaller than the EM wavelength) \cite{Felbacq:2005tm,Rizza:2013em,Ciattoni:2015nh,Rizza:2015oc}, we introduce the small positive parameter $\eta = \tau / \Delta t\ll1$, and the {\em fast} coordinate $T=t/\eta$ (regarded as independent from the {\em slow} one $t$), in order to isolate the slowly and rapidly
varying field contributions \cite{Sanders:1985am}. To ensure finite results in the asymptotic limit $\eta \rightarrow 0$, the basic ansatz of our multiscale approach is \cite{Ciattoni:2015nh} (see also \cite{SuppMat} for details)
\begin{equation}
{\bf D}({\bf r},t)\simeq\overline{\bf D}({\bf r},t)+\eta^2 \widetilde{\bf D} ({\bf r}, t,T),
\label{eq:ansatz}
\end{equation}
Here and henceforth, the overline and tilde denote the slowly and rapidly varying terms, respectively, i.e., $\overline{\bf D}({\bf r},t)={\tilde \tau}^{-1} \int_0^{\tilde \tau} {\bf D} ({\bf r}, t,T) dT$ (with ${\tilde \tau}=\tau/\eta$), and $\widetilde{\bf D}=\eta^{-2}({\bf D}-\overline{\bf D})$. Likewise, to separate the fast and slow contributions in the wave equation, we can decompose the (reciprocal of the) relative-permittivity pulsed waveform as $\varepsilon_p^{-1}=\overline{\varepsilon_p^{-1}}+\widetilde{ \varepsilon_p^{-1}}(T)$ \cite{Footnote1}.  
By substituting the above expressions in the vector wave equation for $\bf D$, after separating
the slowly and rapidly varying contributions and consistently retaining the terms up
to the second order in $\eta$, we obtain the equations \cite{SuppMat}
\begin{subequations}
\begin{eqnarray}
\label{eff1}
&& \frac{\partial^2 {\bf \overline{D}}}{\partial t^2}-  c^2 \nabla^2 \left(\varepsilon_{eff}^{-1} {\bf \overline{D}} -\frac{\gamma}{K^2} \nabla^2 {\bf \overline{D}} \right)  =0,\\
&&  
{\tilde {\bf D}}=-K^{-2} f(T) \nabla^2 \overline{ \bf D},
\label{eq:EMdyn}
\end{eqnarray}
\label{eq:effmod}
\end{subequations}
governing the slow and fast EM dynamics, respectively, within the time interval $0<t<\tau$.
In Eqs. (\ref{eq:effmod}), we have defined $\varepsilon_{eff}=\overline{\varepsilon_p^{-1}}^{-1}$, $K=2 \pi /(c \tau)$, and $\gamma=\eta^2 \overline{\widetilde{ \varepsilon_p^{-1}} f}$, with the function $f$ satisfying the differential equation \cite{SuppMat}
\begin{equation}
\frac{d^2 f }{d T^2} +c^2 K^2 \widetilde{ \varepsilon_p^{-1}}=0,
\label{eq:fT}
\end{equation}
subject to the boundary conditions $\overline{f}=0$ and $\overline{\partial_T f}=0$, which ensure the fulfilling of the  self-consistency constraints $\overline{\tilde {\bf D}}=0$ and $\overline{ \partial_T \tilde {\bf D}}=0$, respectively, since $\tilde {\bf D}$ and $\partial_T \tilde {\bf D}$ are rapidly varying functions \cite{SuppMat}.
We observe that, even though the physical setup is different, Eq. (\ref{eff1}) exhibits the same formal structure encountered in the case of temporally periodic metamaterials \cite{Rizza:2022ne}.
Also in the SPM regime of interest here, the wave-matter interaction is essentially described by an effective relative  permittivity [given by the harmonic average of the waveform $\varepsilon_p(t)$, and accounting for the local response] and a peculiar nonlocal contribution [the term proportional to $\gamma$ in Eq. (\ref{eff1})]. Similar to conventional spatially modulated metamaterials, the rapidly varying contributions can affect the dynamics of the main slowly varying field, yielding strong spatial dispersion and optical magnetism \cite{Ciattoni:2015nh}. 

To study the interaction of an impinging EM wavepacket with the SPM, as for  canonical temporal boundaries \cite{Xiao:2014ra}, we enforce the continuity of the microscopic electric ($\bf D$) and magnetic ($\bf B$) inductions at the temporal boundaries $t_b=0,\tau$. 
By considering plane-wave illumination, ${\bf D}=2 {\rm Re}[ {\bf d}({\bf k}, t) e^{i {\bf k} \cdot {\bf r}}]$ and ${\bf B}=2 {\rm Re}[ {\bf b} ({\bf k},t) e^{i {\bf k} \cdot {\bf r}}]$, and using Eqs. (\ref{eq:effmod}) together with the relevant Maxwell's curl equation, we obtain for the slowly varying terms at $t_b=0$ and $t_b=\tau$
 \cite{SuppMat}
\begin{subequations}
\begin{eqnarray}
\label{bound}
&& \overline{{\bf d}}_{out} =\left(1 + \alpha_0 \frac{k^2}{K^2}\right) \overline{\bf d}_{in},\\
&& \overline{{\bf b}}_{out} = \left(1 + \alpha_0 \frac{k^2}{K^2}\right) \overline{\bf b}_{in} -i c \mu_0 \beta_0  \frac{\bf k}{K} \times \overline{\bf d}_{in},
\end{eqnarray}
\label{eq:BC}
\end{subequations}
where the labels $in$  and $out$ refer to the induction boundary values inside (i.e., $t_b=0^+$ and $\tau^-$) and outside (i.e., $t_b=0^-$ and $\tau^+$) the interval, $\mu_0$ is the vacuum magnetic permeability,  and
\begin{equation}
\label{ab}
\alpha_0(t_b)=\eta^2 f(t_b/\eta),\quad 
\beta_0(t_b)=-\frac{\eta}{c K} \frac{d f(t_b/\eta)}{dT}. 
\end{equation}
Similar to the spatial counterpart (multilayered metamaterials), also in the temporal case the nonlocality affects the interface conditions and it is ruled by fundamental symmetries. A spatial multilayered metamaterial can exhibit strong  nonlocality and chiral boundary effects due to the parity symmetry breaking \cite{Gorlach:2020bc}. In our scenario here, as any time-varying medium, an SPM implies an inherent time-reversal symmetry breaking. By comparison with the conventional (unit-step) temporal-boundary  conditions \cite{Xiao:2014ra} (i.e., the continuity of $\bf d$, $\bf b$ at $t_b$),  it is evident that the nonlocal terms in Eqs. (\ref{eq:BC}) enhance the time-reversal asymmetry. In particular, the term proportional to $\beta_0$ explicitly breaks the time-reversal symmetry. It is worth noting that similar and related effects have also been observed in topological photonic  crystals \cite{Lustig:2018ta,Henriques:2020tp}. Consequently, by tailoring the dielectric pulse profile, it is possible to tune the nonlocality and, in turn, the degree of time-reversal symmetry breaking. This provides new degrees of freedom and enables to fully exploit the time-reversal symmetry breaking for attaining unconventional light-manipulation effects. In order to solve the above boundary-value problem, we need to preliminarily  integrate Eq. (\ref{eq:fT}), so as to compute the effective parameters $\gamma$,  $\alpha_0$ and $\beta_0$. As detailed in \cite{SuppMat}, this can be addressed systematically for a broad class of relative-permittivity pulsed profiles satisfying the condition
$\varepsilon_p(0)=\varepsilon_p(\tau)$, for which
$\alpha_0(0)=\alpha_0(\tau)$, $\beta_0(0)=\beta_0(\tau)$, and therefore a Fourier-series parameterization can be effectively exploited, along the lines of previous studies \cite{Barati:2020tm,Salary:2020tm}.
Under these conditions, by solving Eqs. (\ref{eq:effmod}) with the temporal boundary conditions in Eqs. (\ref{eq:BC}), we obtain the following asymptotic approximation for the  reflection (backward-wave) coefficient \cite{SuppMat}
\begin{equation}
\label{t_r}
r_d(k) \simeq \frac{1}{2} \left( 1- \frac{n_2}{n_1} \right) +i p_1 \frac{k}{K}  +p_2 \frac{k^2}{K^2}, 
\end{equation}
where 
\begin{subequations}
\begin{eqnarray}
p_1&=& \pi \left( \frac{n_2}{\varepsilon_{eff}} - \frac{1}{n_1} \right),\\
p_2&=&  \pi \beta_0 \left( \frac{n_2}{n_1}+1  \right)+\frac{\pi^2}{\varepsilon_{eff}} \left( \frac{n_2}{n_1}-1  \right),
\end{eqnarray}
\label{eq:p1p2}
\end{subequations}
with $k=|{\bf k}|$, $n_j=\sqrt{\varepsilon_j}$ ($j=1,2$). 
Note that the result in Eq. (\ref{t_r}) is obtained in the asymptotic regime $k/K \ll 1$, by neglecting terms of order  ${\cal O}(k^3/K^3)$. By recalling that, for an impinging monochromatic wave with time-dependence $\exp{(-i \omega_1 t)}$,  $\omega_1 = k c/\sqrt{\varepsilon_1}$ and $k/K \sim \omega_1 \tau$, this approximation is fully consistent with our SPM definition \cite{SuppMat}. Likewise, a formally similar approximation can be derived for the transmission (forward-wave) coefficient (see \cite{SuppMat} for details).

Equation (\ref{t_r}) represents a key result of our study, which clearly elucidates the role played by the local effects (i.e., constant term) and dominant nonlocal  contributions (i.e., terms proportional to $k$ and $k^2$). Remarkably, it also suggests simple strategies to tailor their interactions in order to perform elementary wave-based analog computing, which is experiencing a renewed interest within the framework of computational metastructures \cite{Silva:2014pm,Zangeneh-Nejad:2020ac}. For instance, nonlocal effects can be made dominant by assuming the same initial and final media ($\varepsilon_1=\varepsilon_2$), so that the constant (local) term vanishes. Under these conditions, by recalling the derivative property of the Fourier transform, it is apparent that the backward wavepacket contains a linear superposition of the first and second spatial derivatives of the incident one. More specifically, we can identify two parameter regimes of interest. The first one is attained when 
\begin{equation}
\varepsilon_{eff} \neq \varepsilon_2=\varepsilon_1,
\label{eq:regime1}
\end{equation}
so that, in Eq. (\ref{t_r}), the constant term vanishes and the quadratic term is negligible, and thus the backward wavepacket is proportional to the first-order derivative of the incident one. The second regime corresponds to
\begin{equation}
\quad \varepsilon_{eff} = \varepsilon_2=\varepsilon_1, \quad \beta_0 \neq 0,
\label{eq:regime2}
\end{equation} 
i.e., the vanishing of both the constant and linear terms in Eq. (\ref{t_r}), thereby yielding the second-order derivative of the incident wavepacket. It is worth noting that the requirement $\beta_0 \neq 0$ implies that the boundary nonlocality plays a key role in this latter scenario.  In addition, we highlight that the predicted effects are quite robust and not critically dependent on a specific dielectric-permittivity pulsed profile. 

To illustrate the potential of SPMs, we consider several examples, in which the relative permittivity varies between $1$ and $10$. These values are consistent with our underlying assumption of negligible temporal dispersion, and are reasonably feasible in the microwave and terahertz ranges, where high-refractive-index materials are available and can be rapidly modulated in time. As an  example, at terahertz frequencies, by using an infrared femtosecond laser pulse, the dielectric permittivity of a semiconductor (e.g., GaAs, Si) slab can undergo a temporal modulation with significant depth, on a timescale of $\sim$100 fs \cite{Kamaraju:2014sc,Yang:2017tg}.  Therefore, an SPM relying on such a platform should have a duration $\tau \simeq 100$ fs for manipulating a terahertz wavepacket with a characteristic timescale $\Delta t \simeq 10 \tau$=1 ps. Overall, the practical implementation of temporal metamaterials remains very challenging from the technological viewpoint, especially at higher frequencies, although some notable progress is being made (see, e.g., Ref. \onlinecite{Galiffi:2022po} for a recent survey of experimental results).

As a first illustration of SPM, we consider a rectangular pulse, i.e.,  $\varepsilon_p(t)=\varepsilon_{3}$. For this particular profile (conventional temporal slab), an exact closed-form solution is available \cite{Mendonca:2003tb,SuppMat}, not   restricted to the SPM regime $\tau/\Delta t \ll 1$. It is readily verified \cite{SuppMat} that this exact result consistently reduces to  Eq. (\ref{t_r}) (with $\varepsilon_{eff}=\varepsilon_3$ and $\beta_0=0$) for $k/K \ll 1$. Thus, such an SPM can perform the first derivative in the parameter regime of Eq. (\ref{eq:regime1}), but not the second derivative [cf. Eq. (\ref{eq:regime2})], since the intrinsic nonlocality  is absent ($\beta_0=0$).
   
\begin{figure}[t]
\centering
\includegraphics[width=.95\linewidth]{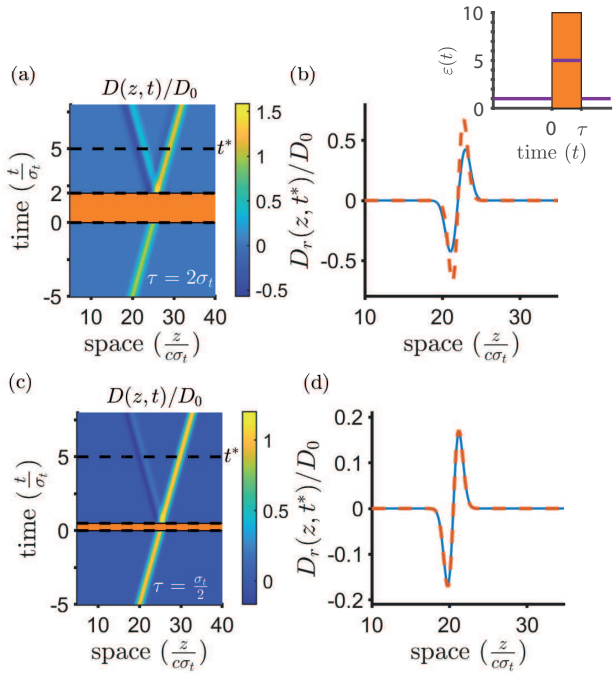}
\caption{Rectangular-pulse profile with $\varepsilon_p(t)=\varepsilon_3=5$, $\varepsilon_1=\varepsilon_2=1$ (temporal slab; see inset). (a) Space-time propagation of a Gaussian wavepacket (normalized electric induction) predicted by the exact theory \cite{Mendonca:2003tb} for $\tau=2 \sigma_t$, with the orange-shaded area representing the temporal slab.
	 (b) Comparison between the backward-wave profiles at time $t^*=5 \sigma_t$ [$D_r(z,t^*)$] predicted by the exact (blue-solid) and SPM (orange-dashed) theories. (c), (d) Same as panels (a), (b), respectively, but for $\tau= \sigma_t/2$.}
\label{Fig2}
\end{figure}

Figure \ref{Fig2} illustrates some representative results, assuming $\varepsilon_1=\varepsilon_2=1$, $\varepsilon_3=5$, and
an impinging Gaussian wavepacket with profile (for $t<0$) 
\begin{equation}
\label{gauss}
{\bf D}_{in}(z,t)=D_0 e^{-\left[\frac{z-v_1 \left(t - t_s \right)}{v_1 \sigma_t} \right]^2} \hat{\bf e}_x,
\end{equation}
where $D_0$ is a constant amplitude, $\hat{\bf e}_x$ is an $x$-directed unit vector, $\sigma_t$ a characteristic timescale, $v_1= c/ \sqrt{\varepsilon_1}$, and $t_s=-25 \sigma_t$. Specifically, Figs. \ref{Fig2}a and \ref{Fig2}b show the space-time map and a spatial cut (at fixed time), respectively, of the normalized electric induction field, for a relative-permittivity rectangular profile (displayed in the inset) with $\tau=2 \sigma_t$, i.e., not satisfying the SPM assumption (see \cite{SuppMat} for details). As a consequence (see Fig. \ref{Fig2}b), the agreement between our theory and the exact solution is not good.  
Conversely, if the rectangular-pulse duration is shortened to $\tau=\sigma_t/2$, as shown in Figs. \ref{Fig2}c and \ref{Fig2}d, the agreement becomes excellent, and the backward-wave profile is proportional to the first-order derivative of the incident wavepacket.   
\begin{figure}[t]
\centering
\includegraphics[width=1.\linewidth]{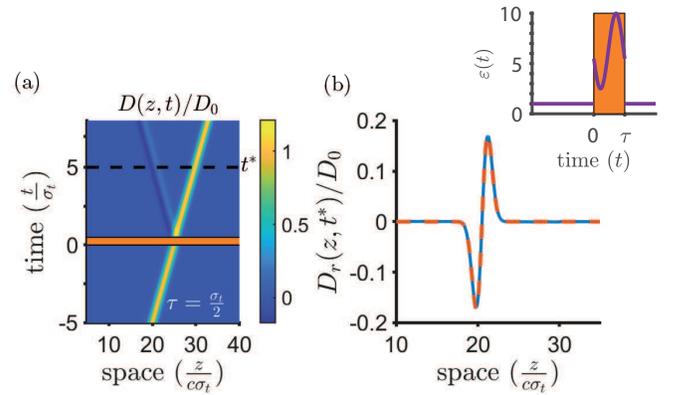}
\caption{Sinusoidal-pulse profile as in Eq. (\ref{cos}), with $\varepsilon_m=6.25$, $\Delta=0.6$, $\phi=\pi$, and  $\varepsilon_1=\varepsilon_2=1$ (see inset).
	(a) Space-time propagation of a Gaussian wavepacket (normalized electric induction) predicted by the SPM theory for $\tau= \sigma_t/2$. 
	(b) Comparison between the backward wave profiles at time $t^*=5 \sigma_t$ [$D_r(z,t^*)$] predicted by full-wave simulations (blue-solid) and SPM theory (orange-dashed).}
\label{Fig3}
\end{figure}
As previously mentioned, for this configuration it is not possible to attain a second derivative. As shown in \cite{SuppMat}, the forward (transmitted) wavepacket is instead dominated by local effects and remains essentially identical to the impinging one. Quite interestingly, this implies that an essentially undistorted copy of the impinging signal is preserved, and can be possibly exploited for further processing.

As a second, more general example, we consider a sinusoidal-pulse relative-permittivity waveform,
\begin{equation}
\label{cos}
\varepsilon_p(t)=\varepsilon_m \left[ 1+ \Delta \cos{\left( \frac{2 \pi}{\tau} t +\phi \right)}  \right], 
\end{equation}   
which naturally satisfies the condition $\varepsilon_p(0)=\varepsilon_p(\tau)$.

\begin{figure}
\centering
\includegraphics[width=\linewidth]{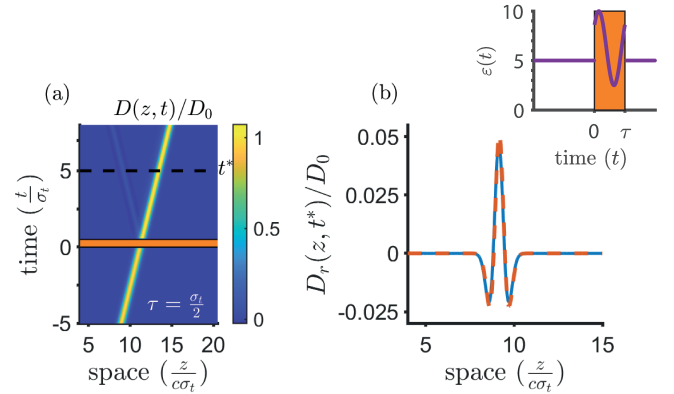}
\caption{As in Fig. \ref{Fig3}, but for $\varepsilon_m=6.25$, $\phi=-\pi/2$, and  $\varepsilon_1=\varepsilon_2=5$.}
\label{Fig4}
\end{figure}
Figures \ref{Fig3}a and \ref{Fig3}b illustrate the interaction with the Gaussian wavepacket in Eq. (\ref{gauss}), assuming
 $\varepsilon_1=\varepsilon_2=1$, $\varepsilon_m=6.25$, $\Delta=0.6$, $\phi=\pi$, and $\tau= \sigma_t/2$. In this case, the  effective relative permittivity is $\varepsilon_{eff} \simeq 5$, and the considered SPM behaves as the temporal slab in the previous example (cf. Figs. \ref{Fig2}c,d). In both cases, the nonlocal parameter $\beta_0$ [and hence the quadratic term in Eq. (\ref{t_r})] vanishes. In fact, from Eqs. (\ref{eq:EMdyn}) and (\ref{ab}), one can show that this effect is general, since it stems from the time-reversal symmetry of the relative-permittivity waveform [$\varepsilon_p(t-t_b)=\varepsilon_p(t_b-t)$]. For the profile in Eq. (\ref{cos}), no exact solution is available, and therefore we validate our predictions against full-wave simulations (see \cite{SuppMat} for details). Figure \ref{Fig3}b compares the electric-induction profiles of the backward (reflected) wave $D_r(z,t^*)$ (at $t^*=5 \sigma_t$) predicted by our SPM theory with the full-wave numerical results, for $\tau=\sigma_t/2$. As for the previous example, the agreement is very good, and a first derivative is attained. Also in this case, as expected, some visible deviations are observed when the duration $\tau$ increases (see \cite{SuppMat} for details). 

As stated above, by tailoring the relative-permittivity pulsed waveform, we can also work in the second regime described by Eqs. (\ref{eq:regime2}). 
This is illustrated in Fig. \ref{Fig4}, where we assume $\varepsilon_1=\varepsilon_2=5$,  $\varepsilon_m=6.25$, $\Delta=0.6$, and $\phi=-\pi/2$, which yield  $\varepsilon_{eff}=\varepsilon_1=\varepsilon_2$, $\beta_0 \simeq 0.13$. In this case, $\varepsilon_b(t)$ does not exhibit even symmetry, so that $\beta_0 \neq 0$; this additional breaking of the time-reversal symmetry enables the computation of the second-order derivative, as can be observed from Fig. \ref{Fig4}b. Once again, the agreement between our theory and the full-wave simulations is excellent in the SPM regime (e.g., $\tau=\sigma_t/2$), and progressively deteriorates as $\tau$  increases (see \cite{SuppMat} for details).

In conclusion, we have studied the hitherto largely unexplored {\em short-pulsed} regime in temporal metamaterials, corresponding to time variations of the dielectric permittivity applied on intervals much shorter than the characteristic wave-dynamical timescale. To this aim, we have developed a general approximate theory, based on a multiscale asymptotic approach, whose predictions agree very well with full-wave numerical simulations, and capture in simple terms the relevant (non)local and symmetry-breaking effects that come into play. As possible application examples of  SPMs, we have illustrated their inherent wave-based analog-computing capabilities, such as the computation of first and second derivatives on an impinging wavepacket. It is worth noting that similar computational functionalities (first-order derivative) were also recently observed in temporal metamaterials characterized by long periodic modulations \cite{Rizza:2022ne}. Our results here indicate that  the necessary nonlocal effects can be equivalently attained, suitably harnessed, and possibly extended (second-order derivative) also in the SPM regime.
The possibility of manipulating EM wavefronts with short-pulsed (rather than long, periodic) temporal variations of the constitutive parameters is conceptually analog to the ``metasurface vs. metamaterial'' juxtaposition in the spatial case, and should be viewed as an elementary brick that opens up new perspectives within the overarching framework of space-time metastructures. Of particular interest for present and future studies are the {\em compound} effects of multiple, time-resolved SPMs
(in analogy with cascaded metasurfaces \cite{Raeker:2019cm}), as well as the joint application of the SPM concept and anisotropy  \cite{Pacheco-Pena:2020ta,Pacheco:2021te,Pacheco-Pena:2021si,Xu:2022gt}. Furthermore, the study of {\em traveling-wave} short-pulsed modulations (e.g., along the lines of \cite{Huidobro:2021ht}) also appears very intriguing. 
Finally, the study of realistic platforms for experimental verification remains crucial for enabling the technological viability of the proposed approach.

\section*{ACKNOWLEDGEMENT}
G. C. and V. G. acknowledge partial support by the University of Sannio under the FRA-2020 program.


%

\end{document}